\documentclass[pra,floatfix,twocolumn]{revtex4}

\usepackage[dvips]{graphicx}%
\usepackage{bm}
\usepackage{color}
\usepackage{amsmath,amssymb}

\newcommand{\ket}[1]{\left |  #1 \right \rangle}
\newcommand{\bra}[1]{ \left \langle #1  \right |}

\newcommand{\ketbra}[2]{\ket{#1}\bra{#2}} 
\def \tr{{\textrm {Tr}}}

\begin{document}

\title{Composability of partially entanglement breaking channels  via entanglement assisted local operations and classical communication}
  
%
\author{Ryo Namiki}\affiliation{Department of Physics, Graduate School of Science, Kyoto University, Kyoto 606-8502, Japan}

\date{\today}
\begin{abstract} 
We consider composability of quantum channels from a limited amount of entanglement via local operations and classical communication (LOCC). 
We show that  any $k$-partially entanglement breaking channel  
  can be composed from    an entangled state  with Schmidt number of $k$   via one-way LOCC.
From the entanglement assisted construction we can reach an alternative definition of partially entanglement breaking channels.

  \end{abstract}

\maketitle

Quantum entanglement is considered to be a resource of quantum information processing. In measurement-based one-way quantum computation any unitary-gate operation can be implemented by sequentially performing local measurements based on classical communication of measurement outcomes once a large entangled state is shared between the nodes \cite{Rauss01,KLM01,G-C99}. An outstanding example is a quantum teleportation gate that transmits the state of an input node to a possibly remote output node by consumption of a maximally entangled state. It establishes an identity quantum channel via local operations and classical communication  (LOCC) with the help of non-local coherence due to entanglement. 
Beside such an LOCC implementation  of unitary gates, 
an interesting research direction would be to identify the role of entanglement for  non-unitary quantum gates or channels. 
 It is always possible to simulate non-unitary gates by deliberately inducing errors on unitary gates. In turn, a direct construction of non-unitary gates naturally addresses the possibility to  save the consumption of entangled resources.

An important class of non-unitary quantum gates in entanglement study is the class of entanglement breaking (EB) channels \cite{{HSR2002}}. Any EB channel has a measure-and-prepare form and can be decomposed into a local measurement and local state preparation based on the classical communication of the measurement outcome. Hence, the class of EB channels constitutes the class of non-unitary gates that can be implemented without entanglement.  As a generalization of EB channels, partially entanglement breaking (PEB) channels have been introduced in Ref. \cite{Chru06} (See also \cite{Hua06}). 
 The notion of PEB channels gives a classification of whole quantum channels based on Schmidt number of entanglement \cite{Ter20}.
  Schmidt number of $k= 1$ identifies the class of separable states, and the corresponding class of  EB channels ($1$-PEB channels) can be composed  solely by LOCC.   On the contrary,  one has to consume an entangled resource in order to construct a PEB channel of Schmidt class $k$ (a $k$-PEB channel) for  $k \ge 2$. It might be fascinating if  the class of $k$-PEB channels constitutes the class of non-unitary gates that can be implemented by entanglement assisted LOCC protocols from entangled resources of Schmidt number $k$.  
    However, the original definition of PEB channels is based on Schmidt number of channel's  isomorphic quantum states given by Choi-Jamiolkowski (CJ) correspondence,
 and has not been connected to the LOCC implementation of quantum channels. 

In the context of quantum benchmark problems \cite{Bra00,namiki12R,Kil10}, the performance of the quantum channels can be estimated by an amount of entanglement of CJ states. It can be related to 
 the amount of entanglement generated by a single use of the channel. To continue the study of quantum benchmarks in quantitative regime  \cite{namiki12R,Kil10}
it might be valuable to investigate the relation between this attainable amount of entanglement due to the channel action and the amount of the resource entanglement to compose the channel by using LOCC.

In this report we  investigate how to implement non-unitary quantum channels via entanglement assisted LOCC.
We show that any  $k$-PEB channel can be constructed from a $k$-dimension maximally entangled state via LOCC. It gives an alternative definition of PEB channels.  Throughout this report we assume quantum channels acting on $d$-level (qudit) quantum systems and $k \le d$.

We begin by giving a physical meaning of CJ correspondence.
Suppose that Alice prepares a maximally  entangled pair $\hat \Phi_{AB}$ 
 and  sends system $B$ to Bob through a quantum channel $\mathcal E$. Then, the final state shared between Alice and Bob $J_\mathcal{E}= id_A \otimes \mathcal{E}_B( \hat  \Phi_{AB}) $ is the CJ state isomorphic to the  channel $\mathcal E$. 
 The amount of possibly shared entanglement $J$ can be associated with the performance of the quantum channel $\mathcal E$ in generalization of the quantum benchmarks \cite{namiki12R,Kil10}.  In this sense, quantitative quantum benchmarks concern the obtainable amount of entanglement by a single use of the quantum channel $\mathcal E$.

\begin{figure}[ptb]
    \includegraphics[width=.8 \linewidth]{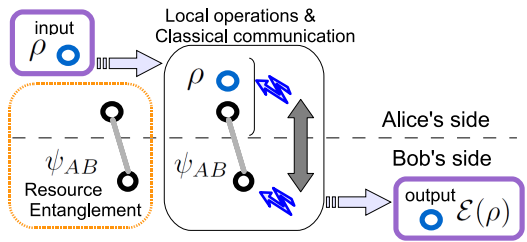}
  \caption{Alice and Bob compose a quantum channel $\mathcal E$ by using a shared entanglement $\psi_{AB}$ with local operations and classical communications (LOCC). An input state $\rho$ is given at Alice's side and the output state $\mathcal E( \rho) $ has to be prepared at Bob's side. 
   }
  \label{fig:SN-fig1.eps}
\end{figure}

Let us consider another situation that Alice and Bob implement the quantum channel by an entanglement assisted LOCC protocol as in FIG. \ref{fig:SN-fig1.eps}. Suppose that an entangled pair $\psi_{AB}$ is shared between Alice and Bob (Alice and Bob possess  systems $A$ and $B$, respectively). They try to compose a quantum channel $\mathcal E$ from the shared entanglement $\psi$ via LOCC provided that an input density  operator $\rho $ is given at Alice's station and the output $\mathcal E (\rho )$ has to be made at Bob's station. If they can establish the channel, it is possible for them to share the entanglement $J$ by transmitting a half of a maximally entangled pair locally prepared by Alice through the channel.  In this procedure, they  start from an initial entanglement $\psi$ and end up with the final entanglement $J$ via LOCC. Due to the monotonicity of entanglement under LOCC, the amount of entanglement to compose the channel has to be no smaller than the amount of entanglement on the channel's isomorphic state $ J $. We thus have a basic relation on the LOCC implementation of the quantum channel  
\begin{eqnarray}
E(\psi ) \ge E(J) \label{choku}
\end{eqnarray} where $E$ is a measure of entanglement (entanglement monotone) \cite{mono}.

Here, we may ask whether this inequality can be saturated. Could it be possible to construct the channel from the entangled resource specified by the isomorphic states? This is true for the case of unitary channels where $J$ is a maximally entangled state and a local unitary action after the quantum teleportation process enables us to implement any unitary channel  from a maximally entangled state via LOCC. 
This is also true for the case of EB channels \cite{HSR2002} where  $J$ is separable and the channel can be implemented without entanglement. 
However, in general, it seems difficult to know how to compose non-unitary channels and to address the relation between $E(\psi)$ and $E(J)$ beyond the general inequality of Eq. (\ref{choku}).  In what follows we consider Schmidt number  \cite{Ter20} as a measure of entanglement  
 and  show that the inequality of Eq. (\ref{choku}) can be saturated by composing a one-way LOCC protocol.

 Before to start, it might be worth to note that  an inverse map of the CJ correspondence is given by $\mathcal E (\rho )= d \tr_A [ (\rho^T\otimes I )  J_{\mathcal E}] $ where 
  $\rho^T$ is transposition of $\rho$  \cite{Nielsen01,Ishizaka08}. In this representation, the input state is transposed, and it could not give us the physical implementation of the channel.   Nevertheless, an entanglement cost for quantum channels is defined through an amount of entanglement to generate a type of CJ state for tensor product of the  channel $\mathcal E^{\otimes n}$ in Ref. \cite{Berta08}. 
Note also that there have been extensive works to quantify the amount of entanglement for an implementation of global unitary operation \cite{C99,Soeda11}.

\begin{table*}[tb]
  \begin{tabular}{l}
      \hline
 \hline
i.   Schmidt number of CJ state is no larger than $k$. \\
     ii. There exists a Kraus representation where the maximum rank of the Kraus operators is no larger than $k$. \\
     iii. The channel can be constructed by an entanglement assisted LOCC from an entangled state with Schmidt number $k$. \\  \hline
 \hline
  \end{tabular}
 \caption{Three equivalent definitions of $k$-PEB channels. The equivalence is proven in the following chain:  i $\Rightarrow$ ii $\Rightarrow$  iii  $\Rightarrow$ i. The original definition  (i) implies (ii) [See. Eq. (\ref{defii})] when we concern the Kraus representation \cite{Chru06}. Our Theorem  of an entanglement assisted LOCC construction leads to   ii  $\Rightarrow$ iii. 
  The  LOCC monotonicity leads to  iii  $\Rightarrow$ i (See main text). 
\label{TableI}}
\end{table*}

Schmidt number of a bipartite density operator $ \sigma$ can be defined as  \cite{Ter20}
\begin{eqnarray}
 \textrm{SN}( \sigma ) := \min_{\{ p_i,\varphi _i \} }\{ \max_i  \textrm{SR}( \hat \psi_i )\} \label{SNs}
\end{eqnarray} where $\{ p_i,\varphi _i \}$ is a decomposition of $\sigma = \sum_i  p_i \hat \varphi _i  $ and  Schmidt rank of a pure state $\hat \varphi$ is the rank of its marginal density operator, i.e.,   $ \textrm{SR} (\hat \varphi) = \textrm{rank}[ \tr_B (\hat \varphi )] $. 
Schmidt number of a quantum channel $\mathcal E$ \cite{Chru06,Hua06} is defined by  Schmidt number of its CJ state: $\textrm{SN} (\mathcal E): = \textrm{SN}(J_{\mathcal E})$.
We call a quantum channel $\mathcal E$ is $k$-partially entanglement breaking ($k$-PEB) if $\textrm{SN} (\mathcal E) \le  k$  \cite{Chru06}.
Let us consider 
the Kraus representation  $\mathcal E ( \rho) = \sum_\alpha K_\alpha \rho K_\alpha^\dagger   $ where the set of Kraus operators $\{  K_\alpha \}  $ fulfills the closure relation $\sum_\alpha K_\alpha ^\dagger  K_\alpha = I$.
 Then, from Eq. (\ref{SNs}) Schmidt number of the quantum channel can be associated with  the rank of the Kraus operators as
\begin{eqnarray}
 \textrm{SN}( \mathcal E ) = \min_{\{K_\alpha  \} }\{ \max_\alpha   \textrm{rank}( K_\alpha   )\}. \label{K-deco}
\end{eqnarray}
Therefore,  if $\textrm{SN} ( \mathcal E ) =k $, there exists a Kraus representation with $\textrm{rank}(K_\alpha )  \le  k $  for all $\alpha $.

It might be instructive to 
 define the set of $k$-PEB channels $\mathcal O_k$ as follows 
\begin{eqnarray}
 \mathcal O_k= \left\{ \mathcal E  \ \Big| \  \mathcal E (\rho ) = \sum_\alpha K_\alpha \rho K_\alpha ^\dagger \  \wedge \  \forall \alpha,\  \textrm{rank} (K_\alpha) \le  k  \right\}  \label{kpeb}. 
\end{eqnarray} Note that this definition includes no explicit  optimization process. We can write 
\begin{eqnarray}    
& \mathcal E  \in  \mathcal O_k  \Leftrightarrow &\nonumber \\
&  \exists \{K_\alpha \} s.t.,  \mathcal E (\rho ) = \sum_\alpha K_\alpha \rho K_\alpha ^\dagger \wedge   \forall \alpha,  \textrm{rank} (K_\alpha) \le k. \ \  \  \ & \label{defii}\end{eqnarray}
From the definition of Eq.  (\ref{kpeb}) we can see the  ``$\Rightarrow $'' direction holds. Conversely, the existence of a representation  $\{K_\alpha \}$ with $\textrm{rank} (K_\alpha) \le k$ directly implies that the ``$\Leftarrow $'' direction is fulfilled. By using Eq. (\ref{kpeb}) the set of EB channels is given by $\mathcal O_1 $ and the set of whole quantum channels for qudit states is denoted by $\mathcal O_d$. We can observe the hierarchy for the classes of non-unitary channels $\mathcal O_1 \subset \mathcal O_2 \subset \cdots \subset \mathcal O_d $. Unitary channels have Schmidt number of $d$ and belong to $\mathcal O_d$. They can preserve entanglement and maintain full-$d$-dimensional coherence. This is in sharp  contrast to the class of EB channels which 
cannot maintain any coherence as they cannot preserve any entanglement. 
 Hence, a higher value of Schmidt number $k$ suggests the capability of maintaining a higher order coherence.  Based on this concept Schmidt-number benchmark enables us to demonstrate the multi-dimensional coherence in quantum channels \cite{namiki12R}. 
 

Now, we proceed our main result to construct PEB channels by an entanglement assisted LOCC protocol. 

\noindent\textbf{Theorem.---} Any quantum channel $\textrm{SN}(\mathcal E) \le   k$ can be constructed by an entanglement assisted one-way LOCC protocol from an entangled  state $\psi $ with $\textrm{SN}(\psi) = k$. 

\noindent\textit{Proof.---} Let us suppose $\mathcal E \in \mathcal O_k$. Let  $\{K_\alpha \}$ be a set of Kraus operators of $\mathcal E$ with $\textrm{rank} (K_\alpha ) \le k$ for any $\alpha$. For any given Kraus representation one can associate an environment system prepared in a certain initial state $\ket{e}_E$ and a global unitary operator $U$ so that $K_  \alpha = \bra{\alpha} U \ket{e}$ holds due to Stinespring dilution, namely, there is  a  decomposition with the environment and global unitary evolution $\mathcal E (\rho) = \tr_E [ U \rho \otimes (\ketbra{e}{e })_E U^\dagger  ] = \sum_\alpha  \bra{\alpha} U \rho \otimes (\ketbra{e}{e })_E U^\dagger  \ket{ \alpha } $, where $\{ \ket{\alpha  }\}$ forms a positive operator valued measure (POVM) on system $E$ and fulfills $\sum_\alpha  \ketbra{\alpha}{\alpha } = I_E$. 
Then, it is possible for Alice to implement the channel action locally by preparing ancilla system with $\ketbra{e}{e}$ and subsequent application of $U$ and the POVM $\{\ket{\alpha }\}$. In this scenario, Alice can conceive which one of the Kraus operators $K_\alpha$ is applied from the measurement outcome of $\alpha$. Since $\textrm{rank} (K_\alpha ) \le k$, Alice can also conceive which one of $k$-dimensional subspaces the conditional state belongs to. 
Hence, the channel output can be transported to Bob's side faithfully by an ideal quantum teleportation of the state in the $k$-dimensional subspace. This can be executed by using a maximally entangled state with Schmidt number of $k$. 
Therefore, Alice and Bob can simulate the single action of  any quantum channel with $\textrm{SN}(\mathcal E)\le k$ by an entanglement assisted one-way LOCC protocol from an entangled state of Schmidt class $k$. \hfill$\blacksquare$

  Note that Alice and Bob can save some amount of entanglement by locally acting the channel on Alice's station. Actually, Alice deliberately induces a decoherence  by performing the POVM, and from the outcome $\alpha $ she can see that a full $d$-dimensional entanglement is not necessary.  An important point is that the rank of Kraus operators determines the amount of entanglement to implement the channel. 
From the construction we can say that the equality in  Eq. (\ref{choku}) is achieved for the case of Schmidt number. 
This implies  that the amount of entanglement to implement a $k$-PEB channel is equivalent to the amount of entanglement of its CJ state as long as Schmidt number is concerned.

We can immediately prove the converse statement of Theorem: Any quantum channel composed from an entangled state of Schmidt number $k$ via LOCC has a Kraus representation in which the maximal rank of the Kraus operators is at most $k$. 
This statement is a direct result of Eq. (\ref{choku}). 
 We can write a formal proof as follows.

\noindent\textit{Proof---} If it is not the case, Schmidt number of the CJ state becomes more than $k$. This contradicts the LOCC monotonicity of entanglement.\hfill$\blacksquare$

Finally, we can identify the class of $k$-PEB channels via entanglement assisted LOCC protocols: 
$k$-PEB channels are quantum channels which can be constructed from an entanglement state of Schmidt class  $k$ via LOCC. We can observe the validity of this novel definition through a simple closed chain between three equivalent definitions  in   Table \ref{TableI}  (Proof of the converse statement constitutes the step iii $\Rightarrow$ i).
The original definition of PEB channels [Table \ref{TableI}(i)] is given by the condition on the CJ state. The second definition [Table \ref{TableI}(ii)] is given by the condition regarding the Kraus form. Our definition [Table \ref{TableI}(iii)] is in a more abstract manner based on a limited amount of entanglement and LOCC. Thereby, we can enjoy three of fundamental aspects on quantum channels to define the class of PEB channels.

 In conclusion we have investigated composability of quantum channels from a limited amount of entanglement via entanglement assisted LOCC.
It has been shown that the class of $k$-PEB channels can be constructed from  entangled resources of Schmidt class $k$ via LOCC ({Theorem}). This gives an alternative definition of PEB channels together with an establishment of a closed chain between different definitions (Table \ref{TableI}).  We hope that the results  offer a basic tool to analyze non-unitary quantum gates and play a key role to find out a general structure on quantum channels.

This work was supported by GCOE Program ``The Next Generation of Physics, Spun from Universality and Emergence'' from MEXT of Japan, and World-Leading Innovative R\&D on Science and Technology (FIRST). 


\begin{thebibliography}{}  
\bibitem{G-C99} D. Gottesman and I. L. Chuang, Nature (London) 402, 390 (1999).
\bibitem{Rauss01} R. Raussendorf and H. J. Briegel, Phys. Rev. Lett. 86, 5188 (2001).
\bibitem{KLM01}E. Knill, R. Laflamme, and G. J. Milburn, Nature (London) 409, 46 (2001). 


\bibitem{HSR2002} M. Horodecki, P. Shor, and M. B. Ruskai, Rev. Math. Phys. 15, 629 (2002).
\bibitem{Chru06}D. Chruscinski and A. Kossakowski, Open Sys. Information Dyn. 13 , 17-26 (2006). 
\bibitem{Hua06} S. Huang, \pra \textbf{73,} 052318 (2006).

 \bibitem{Ter20} B. M. Terhal and P. Horodecki, \pra 61 040301 (2000).



\bibitem{Bra00} 
 K. Hammerer, M. M. Wolf, E. S. Polzik, and J. I. Cirac, \prl \textbf{94,} 150503 (2005); J. Rigas, O. G\"uhne and N. L\"utkenhaus, \pra \textbf{73,} 012341 (2006);  
 R. Namiki, M. Koashi, and N. Imoto, \prl\textbf{101,} 100502 (2008); R. Namiki, \pra 78, 032333 (2008); M. Owari et al., New J. Phys. 10, 113014 (2008); H. H\"aseler, T. Moroder, and N. L\"utkenhaus,  \pra 77, 032303 (2008); J. Calsamiglia et al., Phys. Rev. A 79, 050301(R) (2009); 
 H. H\"aseler and N. L\"utkenhaus,  \pra 80, 042304 (2009); 
 \pra 81, 060306(R) (2010); R. Namiki, \pra 83, 042323 (2011).


 \bibitem{Kil10}N. Killoran, H. H\"aseler, and N. L\"utkenhaus, Phys. Rev. A \textbf{82} 052331 (2010); 
N. Killoran and N. L\"utkenhaus, Phys. Rev. A \textbf{83} 052320 (2011).

\bibitem{namiki12R}R. Namiki and Y. Tokunaga, Phys. Rev. A 85, 010305(R) (2012).


\bibitem{mono}
G. Vidal, J. Mod. Opt. 47, 355 (2000). 

 
 

\bibitem{Nielsen01} M.A. Nielsen and I.L. Chuang,  Phys. Rev. Lett. 79, 321 (1997).\bibitem{Ishizaka08}S. Ishizaka and T. Hiroshima, Phys. Rev. Lett. 101, 240501 (2008). 
\bibitem{Berta08} M. Berta, F.G.S.L. Brandao, M. Christandl, and S. Wehner, arXiv:1108.5357v2. 

\bibitem{C99} J. I. Cirac, A. K. Ekert, S. F. Huelga, and C. Machiavello, Phys. Rev. A 59, 4249 (1999). 
\bibitem{Soeda11}J. Eisert, K. Jacobs, P. Papadopoulos, and M. B. Plenio, Phys. Rev. A 62, 052317 (2000);
D. Stahlke and R. B. Griffiths, Phys. Rev. A 84, 032316 (2011);
  A. Soeda, P.S. Turner, and M. Murao, Phys. Rev. Lett. 107, 180501 (2011).




























\end{thebibliography}
\end{document}